\documentstyle[12pt]{article}

\begin{document}
\begin{titlepage}

\title{Dimensionality, topology, energy, the cosmological
       constant, and signature change
         \thanks{
         Work supported by the Austrian Academy of Sciences
         in the framework of the ''Austrian Programme for
         Advanced Research and Technology''.}}

\author{Franz Embacher\\
        Institut f\"ur Theoretische Physik\\
        Universit\"at Wien\\
        Boltzmanngasse 5\\
        A-1090 Wien\\
        \\
        E-mail: fe@pap.univie.ac.at\\
        \\
        UWThPh-1995-11\\
        gr-qc/9504040}
\date{}

\maketitle

\begin{abstract}
Using the concept of real tunneling configurations (classical
signature change) and nucleation energy, we explore the
consequences of an alternative minimization procedure for
the Euclidean action in multiple-dimensional quantum cosmology.
In both standard Hartle-Hawking type as well as Coleman type
wormhole-based approaches, it is suggested that the action
should be minimized among configurations of equal energy.
In a simplified model, allowing for arbitrary products of
spheres as Euclidean solutions, the favoured space-time
dimension is 4, the global topology of spacelike slices being
${\bf S}^1 \times {\bf S}^2$ (hence predicting a universe
of Kantowski-Sachs type).
There is, however, some freedom for a Kaluza-Klein scenario, in
which case the observed spacelike slices are ${\bf S}^3$. In
this case, the internal space is a product of two-spheres, and
the total space-time dimension is 6, 8, 10 or 12.
\medskip

% PACS-numbers:
\end{abstract}

\end{titlepage}

\section{Introduction}

This paper is devoted to a suggestion concerning the
procedure of minimizing the Euclidean action,
in particular for models in which the cosmological constant
$\Lambda$ is effectively not a fundamental quantity but can
be treated as a degree of freedom by its own. Such models
are encountered
within the standard no-boundary approach of Hartle and
Hawking \cite{HartleHawking} \cite{Hawking2}
as well as in the formalism including
wormeholes and baby universes, due to
Giddings and Strominger
\cite{GiddingsStrominger},
Coleman
\cite{Coleman1} \cite{Coleman2} and others.
Usually, in order to find the dominant contributions to
the underlying path integral,
the Euclidean action (evaluated at solutions
of the field equations) is minimized at constant values
of $\Lambda$ (which yields ${\bf S}^4$ in the
four-dimensional pure gravity case).
In any framework based on the no-boundary proposal, the
minimizing configuration describes the most probable initial
condition for the classical universe at nucleation, and
is associated with the ''probability'' $P\sim\exp(-I)$.
In the approaches including wormholes, the
configurations miminizing the effective Euclidean
action represent
the bunch of universes connected with each other.
Their action determines the measure for the
wormhole parameters, the most famous application thereof
being Coleman's argument that $\Lambda=0$ in all
observations. We will always have these two types of
frameworks in mind, although the relevant objects and
procedures may play a different role therein.
The overall philosophy of these approaches is the same
\cite{Coleman2}. However, an essential technical
difference between them is that the models including
wormholes and baby universes involve approximations
that make them less applicable to quantum cosmological
situations (i.e. to questions concerning the
''history'' of the very early universe, as e.g.
whether it can nucleate with $\Lambda\neq 0$).
In contrast, the (non-wormhole) Hartle-Hawking type models
are formulated in terms of cosmologically relevant
quantities from the outset, and ''nucleation of the
universe'' is quite clearly associated with a wave function
that behaves exponential in some domain of superspace and
oscillatory everywhere else.
Nevertheless, we will formulate our proposal with the
motivation that it might be applicable for both.
\medskip

It has already been observed that admitting
configurations of arbitrary dimensions
leads to the prediction that the observed dimension
is very large. Several modifications have
been suggested in order obtain a
framework that predicts the observed space-time
to be four-dimensional (see e.g. Refs.
\cite{Volovich}--\cite{Myers2}).
Here, we would like to suggest an alternative point of
view, based on a modified concept of ''Euclidean solutions''.
Usually, the action is minimized among solutions
of the Euclidean field equations, each solution (instanton)
being defined on a compact manifold
$\cal M$ without boundary. Our starting point is that
an instanton (sloppily denoted by ${\cal M}$) is not
directly related to physical observables.
On the other hand, the nucleation of the universe as a
tunneling phenomenon can be described
in terms of a ''real tunneling geometry'', i.e.
a classical field configuration whose metric signature
changes from Euclidean to Lorentzian type across a
hypersurface $\Sigma$, which plays the role of the
initial spacelike slice of a classical universe
\cite{GibbonsHartle}. The Euclidean part of the
corresponding manifold represents
the ''tunneling from nothing'', whereas ''at''
$\Sigma$ the classical (Lorentzian) time emerges.
This situation can be encoded in terms of a purely
Euclidean configuration, defined on $\cal M$,
together with a hypersurface of zero extrinsic
curvature (or more general: a ''symmetric''
hypersurface).
These issues are reviewed in Section 2.
Our proposal amounts to replace ${\cal M}$
by the pair $({\cal M},\Sigma)$ as the class of
objects to be compared by means of their
Euclidean action. (In a wormhole-based framework, the
interpretation of $\Sigma$ is less straightforward than
in (non-wormhole) Hartle-Hawking type quantum cosmology,
but at least formally it provides an
$(n-1)$-dimensional ''space'' $\Sigma$ associated with an
$n$-dimensional instanton ${\cal M}$).
Interpreting $\Lambda$ as a vacuum energy
density, we associate with any such pair
$({\cal M},\Sigma)$ a nucleation energy
$E\sim \Lambda \times$ ${\rm volume\,\,of\,\,\Sigma}$
(including additional matter terms in general). It
is then suggested that -- as an alternative to the
standard procedure -- the minimization of the Euclidean
action could be performed at constant $E$. This scheme
enables one to extract probabilities for (or at least a
relative order among) the real tunneling configurations
in different dimensions.
As a consequence, low
dimensions (large energies) are favoured, and the most
probable topology of spacelike slices may be predicted
even in multiple-dimensional theories.
In Section 3, we formulate our proposal and give
some heuristic motivations.
We should emphasize here that all approaches based on the
value of the action on solutions of the Euclidean field
equations (i.e. the ''on shell''-action)
provide only approximations to the full
quantum theory for certain types of questions, the
procedure we suggest being included of course.
What we have in mind in the present paper are questions
concerning the ''large-scale'' variables, in particular
the dimension and topology of space and space-time.
Despite the inclusion of ''effective''
phenomena (such as the cosmological constant), the
underlying small-scale dynamics has largely been truncated,
so that e.g. the issue of decoherence
\cite{GellMannHartle}--\cite{Hartle} is
inaccessible within such a framework. Neverthless,
it is just this issue that accounts for
the selection of a real, classical universe. We
understand our suggestion as a possible effective
imprint of the ''true'' underlying quantum theory
on the approximate level of a minimize-the-action
procedure.
\medskip

In order to have a computable scenario at hand, we present
a somewhat simplified model in Section 4, neglecting
all matter contributions except for the cosmological
constant, and admitting arbitrary products of
spheres in any dimension as the fundamental configurations
${\cal M}$. The associated hypersurfaces $\Sigma$ are
provided by the equators of the factor spheres. The
first step in minimizing the action, keeping the energy
$E$ fixed as well as the dimension $n$, is carried out
in Section 5. We encounter
a sequence of minimizing configurations, denoted by
${\cal K}_n$, each being a product of two-spheres with
some $p$-sphere ($p\sim\sqrt{n}$ for large $n$).
It turns out that for large $E$ the
favoured dimension is $n=4$, the corresponding
spacelike slice $\Sigma$ being ${\bf S}^1 \times {\bf S}^2$,
i.e. of Kantowski-Sachs type.
If the minimization is only approximate (caused by
the details of some decoherence mechanism or by an
effect at the wormhole scale), there is some
freedom for a Kaluza-Klein scenario with
spacelike slice $\Sigma={\bf S}^3$ and total dimension
$n=6$, $8$, $10$ or $12$. The internal (compactified) space
is a product of two-spheres.
These results provide the main arguments
in favour of our proposal.
In addition, we discuss the configurations at low energies,
and find evidence for a higher-dimensional phase of the
universe ''before'' nucleation.
\medskip

In Section 6, we demonstrate the (appearently quite generic)
feature that minimization at fixed but large $E$ leads to
low dimensions, whereas in Section 7 we show that
among the configurations selected already, those with
large $E$ are favoured. This indicates that
nucleation can be expected at large energies, and
provides the completion of the arguments
leading to the prediction of the
above-mentioned configurations.
The role of $E$ being large at nucleation is formally
related to the smallness of $\Lambda$ in the
standard wormhole theory. Thus we mention that our
framework might be of some significance for the late
(classical) universe as well, $(E,\Sigma)$ representing
the actual energy and the spacelike slices. If this is
true, our proposal might anticipate some generalization of
Coleman's wormhole model for the case of a
cosmological scenario (and probably the inclusion of
some large-scale decoherence mechanism). We point out that
in a somewhat effective sense our action for the
late universe coincides with Coleman's
(equation (\ref{zsfdssdg}) below).
In Section 8, we
speculate about a modification of the
fundamental gravity action by boundary terms that could
render the real tunneling configurations the basic objects
without imposing any {\it ad hoc} hypothesis.
Finally, we mention the generalization of an alternative
minimization procedure within our framework,
based on a related idea originally due to Gasperini
\cite{Gasperini}, using a different concept of ''energy''.
It predicts space-time to be
${\bf R}\times {\bf S}^3$, although its applicability might
be limited to (products of) spheres.
\medskip

To summarize, the most direct way of extracting a
favoured classical space-time from our model gives
${\bf R}\times{\bf S}^1\times {\bf S}^2$
in $n=4$, but it is also consistent with
${\bf R}\times{\bf S}^3\times {\rm internal\,\,space}$
in $n=6$, $8$, $10$ or $12$. The computations
leading to these results are mainly
contained it Section 5.
(We should add that it is presumably quite easy to write
down a model containing a slightly more sophisticated
matter sector, such that
the favoured classical space-time
is ${\bf R}\times {\bf S}^3$ -- which is certainly
the most appealing configuration).
Thus, from the
point of view of dimensionality, our proposal
reproduces a reasonable behaviour of the early universe.
We must leave it open whether, apart from this issue,
it might as well be helpful with regard to the conceptual
problems of quantum cosmology.
\medskip

\section{Euclidean action and real tunneling configurations}
\setcounter{equation}{0}

One of the drawbacks in Euclidean quantum gravity is the
fact that the Euclidean action is not bounded from below.
Thus, a path integral over ''all'' compact Euclidean (i.e.
positive signature) metrics diverges badly. There have been
techniques developed to impose a separate analytic
continuation in the conformal freedom
\cite{HawkingQG}--\cite{Hawking1}.
Within such a scheme, the dominant contributions in the path
integral are provided by the {\it stationary} points of the
action (i.e. solutions to the Euclidean field equations)
that are not necessarily local minima.
The according procedure consists of evaluating
the Euclidean action ''on shell''. Thus restricted,
the gravitational action (including
a fixed positive cosmological constant) is bounded from
below \cite{Besse}, and the particular solution that
{\it minimizes} the action is expected to be the dominating
one.
\medskip

The Euclidean action in an abitray number of dimensions reads
\begin{equation}
I = -\, \frac{1}{16 \pi G_n}
\int_{\cal M} d^n x\,\sqrt{g}\,\left(R-2\Lambda\right)
- \,\frac{1}{8 \pi G_n} \int_{\partial {\cal M}}
   d^{n-1} x\, \sqrt{h} \,K
+ I^{\rm\, else}\, ,
\label{2.1}
\end{equation}
where $G_n$ is the $n$-dimensional gravitational constant,
and $I^{\rm\, else}$ stands for contributions from matter
fields (and possibly higher order curvature terms).
An action of this type occurs in different frameworks,
either representing a fundamental theory including
various matter interactions, or just as an effective
action obtained after (some or all) non-gravitational
degrees of freedom have been integrated out. If
the possibility of wormholes and topology
fluctuations at small scales is included, the coupling
constants (in particular the cosmological constant $\Lambda$)
depend on the wormhole parameters and may in some sense
be treated as degrees of freedom by their own.
The stationary points of $I$ are the solutions of the
Euclidean Einstein equations
\begin{equation}
R_{\mu\nu} - \frac{1}{2}\, g_{\mu\nu} \,R =
  - \Lambda\, g_{\mu\nu} + 8 \pi G_n\,T^{\rm\, else}_{\mu\nu}
\equiv 8 \pi G_n\,T^{\rm tot}_{\mu\nu}\, .
\label{2.2}
\end{equation}
If $I$ explicitly contains matter fields $\Phi$,
their corresponding equations of motion (obtained from
$\delta I=0$ along with (\ref{2.2})), are taken into account
as well.
Assuming $\Lambda>0$ and a reasonable behaviour of the
matter sector, there are two different types of solutions
$({\cal M},g,\Phi)$ of interest. Either
${\cal M}$ is a compact manifold without boundary,
(in which case we call the configuration an ''instanton'')
or ${\cal M}$ is compact and has a boundary ${\cal N}$ at
which $(h,\phi)$ -- the appropriate restrictions of
$(g,\Phi)$ -- are prescribed.
The instanton configurations of the effective action
arise quite directly in the
evaluation of the path integral when wormholes and baby
universes are included
\cite{GiddingsStrominger} \cite{Coleman1}
\cite{Coleman2}),
whereas the second type of solutions is due to the
conventional situation in quantum cosmology,
when the wave function of the universe $\psi[h,\phi]$ is
computed in the semiclassical approximation
\cite{HartleHawking} \cite{Hawking2}.
For pure gravity ($T_{\mu\nu}^{\rm\, else}=0$),
the most symmetric instanton is the
round sphere ${\bf S}^n$ with radius $a$ given by
$a^2=(n-1)(n-2)/2\Lambda$. The action thereof
provides a general lower bound, i.e. it is less than
$I({\cal M},g)$ evaluated at any other pure gravity
instanton in the same dimension and with the same $\Lambda$
\cite{Besse}.
\medskip

With any instanton configuration $({\cal M},g,\Phi)$, one
usually associates a weight $\exp(-I)$. This is often
treated as if it was an amplitude or probability
(see e.g. Refs. \cite{Volovich}, \cite{Baum}
and \cite{HawkingCC}).
In the path integral approach to
quantum cosmology based on the no-boundary
proposal by Hartle and Hawking
\cite{HartleHawking} \cite{Hawking2}
(which is just the formalization of the requirement that
only compact metrics are involved)
the dominant contribution to $\psi[h,\phi]$ is not
provided by instantons but by solutions with boundary.
However, the instantons describe the nucleation
(''tunneling from nothing'') of the
universe. In the simplest ($n=4$ FRW) minisuperspace example
with $T_{\mu\nu}^{\rm\, else}=0$ and $\Lambda$ fixed,
the only instanton is the four-sphere with radius $a$
($a^2=3/\Lambda$). Any three-sphere with radius $b<a$, when
placed into the instanton manifold as a hypersurface,
divides ${\bf S}^4$ into two pieces, the
action $I_{HH}(b)$ of the smaller one giving rise
to the (approximate) wave function
$\psi(b)\sim\exp(-I_{HH}(b))$.
For $b>a$, the condition that ${\bf S}^4$ contains a
larger ${\bf S}^3$ can only be met by analytic
continuation to complex numbers,
and $\psi(b)$ becomes oscillating, i.e. provides
the semiclassical wave function for the evolution
of the (Lorentzian) universe \cite{HartleHawking}.
Thus, the universe nucleates (becomes ''real'' or
classical) at a scale factor $b\approx a$. Since
$\psi(a)\sim\exp(-I/2)$, where $I$ is the total Euclidean
action of the instanton ${\bf S}^4$, the quantity
$P\sim\exp(-I)$ may be referred to as the
(approximate) ''nucleation probability''.
In similar situations, but with $\Lambda$ being only an
effective vacuum energy density (e.g. induced by a
scalar field, $\Lambda=8 \pi G_n V(\phi)$, in the
approximation that
the dynamics of $\phi$ is ignored), the
nucleation probabilities according to different values
of $\Lambda$ (or to even to different wave
functions) are directly compared, and supposed to
yield an estimate for the most probable classical
initial configuration (see e.g. Refs.
\cite{Vilenkin4}, \cite{GibbonsHartle} and \cite{Lukas}).
A more general concept of the class of all instantons which
provide a ''nucleation seed'' for the universe within
the no-boundary proposal is provided
by the ''real tunneling configurations'' (see below).
\medskip

In a model including wormholes ($I$ denoting the
effective action), the weight $\exp(-I)$ becomes
once more exponentiated, giving rise to a contribution
$\exp(-I)\, \exp(\exp(-I))$ to the measure for the wormhole
parameters \cite{Coleman2}.
The cosmological constant thereby effectively represents
the small scale topology fluctuations
(cf. Ref. \cite{HawkingSF}).
The minimizing instanton configuration
provides the dominant (in some situations even
the {\it only})
contribution to the measure for the wormhole parameters.
This is the basis for Coleman's argument that the
connection of our universe with other universes by
wormholes forces the obverved $\Lambda$ to be
exactly zero (as soon as the universe is larger than the
wormhole scale). The technical reason is that for
fixed $\Lambda$ the action is minimized by the four-sphere
and takes the value $I=- 3\pi/G \Lambda$, which in turn
becomes $-\infty$ as $\Lambda\downarrow 0$.
Originally, Coleman supposed the integration over all matter
fields to have been carried out already
\cite{Coleman2}, although later authors kept some or all
non-gravitational degrees of freedom in the effective action
(see e.g. Refs. \cite{Volovich} and \cite{Myers2}), in
particular if non-trivial matter vacuum configurations exist.
In any case, the action is of the type
(\ref{2.1}), possibly with $I^{\rm\, else}$ being disregarded
or containing only higher order curvature terms.
Minimization of the action at constant $\Lambda$ again gives
(for $n=4$ and pure gravity degrees of freedom contained in
$I$) the instanton ${\bf S}^4$, and only thereafter
the dependence of $\Lambda$ on the wormhole parameters is
taken into
account (leading to the $\delta(\Lambda)$-peak in turn).
However, alternative minimalization procedures have
been suggested.
In particular, if arbitrary dimensions are admitted
by the underlying theory, the minimization of $I$ at
constant $\Lambda$ leads to $n\rightarrow\infty$.
Gasperini \cite{Gasperini} considered
$n$-spheres as instantons in arbitrary dimensions
and proposed that the cosmological constant
depends on the dimension $n$. He minimized the action
at constant values of the
(formal) energy $U$ contained in the
Euclidean geometry on the $n$-sphere
(it is defined by $\Lambda/8\pi G_n$ times the volume
of a ball in ${\bf R}^{n-1}$ with the same radius as
the instanton ${\bf S}^n$,
and is proportional to
$\partial I/\partial\beta$
as formally computed in the thermodynamic approach to
Euclidean quantum gravity; in Section 8 we comment on
this approach in more detail).
His conclusion that flat space-time is
four-dimensional is appealing, but the relation $U$
to an observed energy is not clear.
Other approaches to the dimensionality problem
based on the Euclidean action
may be found in Refs. \cite{GasperiniUmmarino},
\cite{Volovich} and \cite{Myers2}.
\medskip

In what follows we envisage models whose $\Lambda$
is enough of an effective phenomenon such that it
can be treated as an independent variable when the
action is minimized. In the
wormhole formalism, this is obviously the case, but
one may encounter similar situations within
the Hartle-Hawking framework as well (e.g.
if $\Lambda=8 \pi G_n V(\phi)$ or
if some underlying dependence of $\Lambda$
on additional variables -- with or without dynamics --
is assumed; cf. Refs. \cite{Baum} and \cite{HawkingCC}).
One may also consider a third kind of models
-- ranging somewhere in between the Hartle-Hawking and
the Coleman type appoaches -- by naively assuming that
$\Lambda$ effectively encodes the small-scale ''foam''-like
structure of space-time, but without admitting
wormholes that connect different universes
\cite{HawkingSF}. This just renders $\Lambda$
an own degree of freedom as well. With regard to
to conceptual questions, this type of models is closer
related to the Hartle-Hawking formalism, into
which it shall henceforth be included.
Traditionally, as mentioned, in all these cases
the minimization of $I$ (e.g. with respect to topology)
is first carried out at constant
$\Lambda$ (see e.g. Refs. \cite{Coleman2} and
\cite{GibbonsHartle}). If we parametrize the general
instanton solution to a given $\Lambda$ symbolically by
$\gamma$ (which denotes a collection of a huge number of
parameters to characterize the solutions, including
their dimension and topology),
the action of arbitrary instantons
may be written as $I_\Lambda(\gamma)$. Varying with
respect to $\gamma$ at constant $\Lambda$ yields a
minimizing configuration, denoted by $\gamma(\Lambda)$.
(If $n=4$ is kept fixed from the outset, $\gamma(\Lambda)$
just denotes the instanton ${\bf S}^4$ with
$a^2=3/\Lambda$). In a second step,
one associates with $\Lambda$ the number
$I_\Lambda(\gamma(\Lambda))$ which can be minimized
again (to give $\Lambda\downarrow 0$ in general).
\medskip

The aim of this paper is to suggest a modified way of
performing the minimization of $I$. The role of instantons
as ''nucleation seeds'' in the
no-boundary framework has already been mentioned
for the case of the FRW minisuperspace
model. The initial classical spacelike
slice is provided by the three-sphere with radius $a$,
hence the equator of the instanton ${\cal M}={\bf S}^4$. Let
us denote this hypersurface by $\Sigma$. In general,
there is a variety of pairs $({\cal M},\Sigma)$, each
being a candidate for inducing nucleation
(''at'' $\Sigma$ as the initial spacelike slice)
and each being associated with the (approximate) nucleation
probability $P\sim\exp(-I)$.
\medskip

The idea of associating with a Euclidean configuration
$\cal M$ a hypersurface $\Sigma$ can be made more precise by
introducing the concept of ''real tunneling geometries''
\cite{GibbonsHartle}
(or ''real tunneling configurations'', to be more
general) and ''classical signature change''
( see Refs. \cite{Hayward1}--\cite{FE4}).
In analogy with the usual way one
treats tunneling phenomena in quantum mechanics,
the process of ''tunneling from nothing'' of the universe
can be described by a {\it Euclidean} solution of the field
equations that matches smoothly the subsequent classical
evolution in {\it Lorentzian} time. The corresponding
construction is a configuration
$(g,\Phi)$ on some manifold
${\cal M}_{\rm real\,\, tunneling}$
such that across a hypersurface $\Sigma$ the signature of the
metric changes from Euclidean $(+,+\dots,+)$ to
Lorentzian $(-,+\dots,+)$ type .
$\Sigma$ is spacelike with respect to the Lorentzian
side. The Euclidean part ${\cal M}_{\rm Eucl}$ is compact
(its boundary of course being $\Sigma$),
and the field equations (including Einstein's equations,
which are defined independent of the metric signature) are
satisfied in the two single-signature parts of
${\cal M}_{\rm real\,\, tunneling}$.
Moreover, a set of junction conditions implies
that ''the field equations'' are satified ''at'' the changing
hypersurface $\Sigma$ as well:
These conditions require the configuration as a whole to
be a stationary point of the action (which is in this
context considered
as a genuine complex quantity, evaluated at genuine
complex field configurations
\cite{GibbonsHartle}).
They essentially state that the extrinsic curvature induced
by the metric $g$ on $\Sigma$, along with the normal
(affine parameter) derivatives of certain matter
field components must vanish when computed with respect to
either (Lorentzian or Euclidean) side
(see e.g. Refs. \cite{GibbonsHartle} and \cite{Hayward1}).
In other words, $\Sigma$ must carry ''time-symmetric''
initial-data.
In terms of a slicing such that $\Sigma$ is described by
$t=0$ and the Euclidean part ${\cal M}_{\rm Eucl}$ of
${\cal M}_{\rm real\,\, tunneling}$ is given by $t<0$,
the junction conditions imply the following:
If the Euclidean ''time'' evolution starts from some value
$t<0$ and is re-performed, but {\it without} changing the
signature at $t=0$, then the Euclidean solution emerging for
$t>0$ is just the time-reversed (reflected) version of the
solution in the region $t<0$, the total result
being a purely Euclidean solution on a compact manifold
$\cal M$ without boundary (hence an instanton),
carrying a ''time''-symmetric hypersurface $\Sigma$
(In the notation of Ref. \cite{GibbonsHartle},
${\cal M}$ is written as $2 {\cal M}_{\rm Eucl}$).
The quotation marks indicate that $t$ is the Euclidean
time on $\cal M$. We will simply call $\Sigma$
a {\it symmetric} hypersurface.
Likewise, there is a purely Lorentzian
solution with a hypersurface $\Sigma$ carrying
time-symmetric data. The manifold
${\cal M}_{\rm real\,\, tunneling}$
then just consists of half of $\cal M$, joined along $\Sigma$
to half of the corresponding Lorentzian manifold.
If the above-mentioned ''time'' coordinate $t$ is chosen
appropriately, the two parts of
${\cal M}_{\rm real\,\, tunneling}$ are related by a Wick
rotation in $t$
\cite{GibbonsHartle}.
This representation of the ''brith of the universe''
may be referred to as {\it strong} signature
change and should not be confused with {\it weak}
signature change, which is an attempt to
define an alternative theory of gravity \cite{FE4}.
\medskip

To summarize, a real tunneling configuration is provided
by an instanton $({\cal M},g,\Phi)$, together with a
symmetric hypersurface $\Sigma$. Any such
configuration describes a possible nucleation
of the universe, where $\Sigma$ and its data $(h,\phi)$
provide the initial conditions for the subsequent
(classical) time evolution \cite{GibbonsHartle}.
The Euclidean action evaluated over
the tunneling part is $I/2$, where $I$ is the total
action of the instanton (the boundary term in
(\ref{2.1}) being obsolet just because the extrinsic
curvature on $\Sigma$ vanishes).
The according nucleation probability is $P\sim\exp(-I)$.
\medskip

The simplest example of a signature changing solution
of the pure Einstein equations with a positive
cosmological constant $\Lambda$ is provided by
cutting an $n$-sphere along its
equator and joining it to the corresponding half of a
de Sitter space ${\bf H}^n$
(cf. Refs. \cite{Vilenkin1} and \cite{GibbonsHartle}).
The ''radius'' is related to the cosmological constant
by $a^2=(n-1)(n-2)/2\Lambda$, and the real tunneling
geometry is given by the metric
\begin{equation}
ds^2 = -\,{\rm sgn}(t) dt^2 + b(t)^2 d\sigma_{n-1}^2
\label{signch}
\end{equation}
where $t$ is ranging from $-a \pi/2$ to $\infty$, and
$d\sigma_p^2$ is the round metric on the unit $p$-sphere.
Furthermore,
\begin{equation}
b(t) = \left\{ \begin{array}{ccc}
a \cos(t/a)&{}& \qquad\qquad
     - a \pi/2 < t < 0\\
a \cosh(t/a)&{}& \qquad\qquad 0<t\, .
                 \end{array}
        \right.
\label{jdshcjkfn}
\end{equation}
Let us denote the space thus defined by ${\bf K}^n$.
The hypersurface
$\Sigma = {\bf S}^{n-1}$ is given by $t=0$. The region
$t<0$ is half of ${\bf S}^n$, the region $t>0$ is half of
${\bf H}^n$. This configuration represents
the quantum tunneling process occuring in the
$n=4$ FRW minisuperspace model.
The genuine quantum state of the universe (i.e. the state
in which it is in the classically forbidden region
''smaller than the minimum de Sitter radius $a$'')
is visualized as the Euclidean part, whereas
the Lorentzian part represents the subsequent classical
evolution.
\medskip

The physical significance of $\Sigma$ as the initial
classical hypersurface is rather clear if the
underlying concept is the no-boundary wave function
of the universe. However, we would like consider the
identical construction in the
case of the effective action appearing in wormhole-based
models as well. There, the relation
between Euclidean (instanton) configuration and
observable quantities is less direct. In some formal sense,
the instantons are associated with a variety of
universes, connected with each other by wormholes.
However, the precise role played by these configurations
depends on the issue of how to regularize and compute the
path integral (cf. e.g. Ref.
\cite{KlebanovSusskindBanks}), and
thouch upon deeper conceptual questions of quantum gravity.
It is in particular the appearance of arbitrarily large
universes that causes the need for a cutoff
\cite{Coleman2}, and that might look unappealing
within the context of cosmology.
It is conceivable that in some quantum cosmological
modification of Coleman's theory (taking into account the
''smallness'' of the universe, or universes) the instantons
of the effective action play a role comparable to the
Euclidean configurations dominating the wave function
in the no-boundary approach and giving rise to the
concept of nucleation. Thus, we consider the pairs
$({\cal M},\Sigma)$
 -- rather than ${\cal M}$ alone -- as the
relevant objects in this case as well. We admit that
it is not obvious whether such a modification would apply
only to the cosmological context, and whether it
might be relevant for {\it today's} (large) universe
as well (e.g. in the sense that a scale factor $a$ of
$\Sigma$ just represents a quantum contribution
to the actual scale factor observed;
we will speculate
on this at the end of Section 7).
\medskip

Having adopted the concept of real tunneling
configurations $({\cal M},\Sigma,g,\Phi)$ as relevant
quantities in models described by an action of the
type (\ref{2.1}),
we should state that an instanton may admit several
non-equivalent symmetric hypersurfaces $\Sigma$.
On the other hand,
not any $(\Sigma,h,\phi)$ will be associated with
an instanton. Hence, in our symbolic notation,
the set of all ${\cal M}$ is not identical to the set of
all $({\cal M},\Sigma)$. However, we expect all dominant
configurations to be contained in both classes, and
different situations to be suppressed by the path intgral.
\medskip

\section{Nucleation energy}
\setcounter{equation}{0}

We are now in a position to develop our proposal.
If $\Lambda$ is thought of as ''the vacuum energy density'',
one might like to know whether this quantity refers to some
''space'', in which a corresponding amount of energy is
distributed. We follow our line of reasoning and treat
the real tunneling configurations as the relevant ones.
In view of the nucleation scenarios,
it is natural to associate the ''energy density'' referred to
as $\Lambda$ (plus possible contributions from the matter
fields $\Phi$) with the initial classical hypersurface $\Sigma$.
Let us denote by $E$ the total
energy contained in the matter fields (including $\Lambda$)
thus defined. Its density is given by
\begin{equation}
\rho = \frac{\Lambda}{8\pi G_n} -
T^{\rm\, else}_{\mu\nu} \,n^\mu n^\nu
\equiv
-\,T^{\rm tot}_{\mu\nu} \,n^\mu n^\nu
\label{wxnlnbk}
\end{equation}
where $n^\mu$ is the unit normal to $\Sigma$. (In the
formula above, the energy momenum tensor is the Euclidean
one, and $n^\mu n_\mu=1$, hence the unusual sign. However,
this expression has the advantage of being defined with
respect to the Euclidean configuration only).
The total energy associated with this density is thus
\begin{equation}
E = \int_\Sigma d^{n-1}x\,\sqrt{h}\,\rho
\equiv \frac{1}{8\pi G_n}\, {\cal V}_{\Sigma}\, \Lambda
- \int_\Sigma d^{n-1}x\,\sqrt{h}\,
T^{\rm\, else}_{\mu\nu} \,n^\mu n^\nu\, ,
\label{energy}
\end{equation}
where ${\cal V}_\Sigma$ is the volume of $\Sigma$
as an $(n-1)$-manifold.
\medskip

If $\Lambda$ is allowed to take arbitrary values,
(which is our assumption), this last equation can
be solved for $\Lambda$ in terms of $E$ (and the additional
variables, encoding all further properties,
including the dimension and topology). Let us symbolically
denote by $\alpha$ the variables characterizing the
field configuration on $({\cal M},\Sigma)$ at fixed $E$.
Thus, with
each real tunneling configuration we associate
the corresponding value for $E$, and one arrives at
an alternative expression for the action, denoted
by $I_E(\alpha)$. This might be viewed as a trivial
change of variables, but we should keep in mind that
the equation relating $\Lambda$ and $E$ involves all
the variables $\alpha$. (The variables $\alpha$ and
$\gamma$ partially coincide, but are not
identical).
\medskip

Let us proceed heuristically. Nucleation (i.e. the
emergence of a superposition of semiclassical
universes) is probably related to decoherence
\cite{GellMannHartle}--\cite{Hartle}
(i.e. the actual selection of {\it one} universe
that is classical at least with respect to the large-scale
variables such as dimension and topology). We do
not know the details of this selection process.
(Clearly, it cannot be understood in terms of a
formalism based mainly on the value of the action
for instantons and real tunneling
configurations. As a consequence, decoherence
shows up effectively as a process that might be
dealt with heuristically and formally, but
whose origin is external to our
framework). It should
not be a surprise if the emergence of classical time is
related fundamentally to its conjugate quantity, energy.
Minimization of the action is certainly one basic
principle to characterize this mechanism. However, the
action usually does not admit a well-defined minimizing
configuration as long as all its variables are allowed
to vary arbitrarily.
One possibility to guess a futher principle
accoding to which the selection operates
is to conjecture that the ''competition'' among various
configurations favours those at equal energy
prior to those at different energy.
\medskip

One may look at this in an even more heuristic way:
There is some amount of energy $E$ associated with
the initial field configuration on $\Sigma$.
Gasperini \cite{Gasperini} mentioned the
idea that there could have been only a certain amount
of energy ''available'' for the formation of
geometrical structures.
However, in a sense, the total energy
of a closed universe is zero.
Hence, $E$ is precisely compensated by an amount of
energy $-E$ contained in the gravitational field.
Thus one can look at the mechanism that drives the
universe classical as a process with
energy balance $0\rightarrow E - E$. If this
transition of ''zero energy'' into two precisely
cancelling contributions is somehow constrained in a
quantum state, the sub-spaces at constant $E$ in
the space of all configurations
attain a preferred role.
Likewise, one might think about
$E$ to be ''conserved'' during transitions
between configurations with different dimension and
topology.
\medskip

Thus, in technical terms, we suggest to
minimize the action $I_E(\alpha)$
at fixed $E$ (instead of fixed $\Lambda$), thereby
obtaining a set of minimizing
configurations, symbolically denoted by $\alpha(E)$.
The subsequent minimization of $I_E(\alpha(E))$
(in general leading to $E\rightarrow\infty$,
as a sort of analogy to $\Lambda\downarrow 0$)
indicates that the unicerse becomes classical at large $E$.
We do not know exactly at which value of $E$ this happens,
but in the simplified model to be considered in
the remaining Sections, we will encounter a good
candidate. Moreover, the configurations for
large $E$ will turn out to be four-dimensional.
In contrast, the configurations $\alpha(E)$ for small
$E$ have large $n$ and represent the genuine
multiple-dimensional quantum state of the early
universe.
\medskip

As a consequence of the procedure we suggest,
$\Lambda$ effectively
becomes dependent not only on the
dimension (as was anticipated before \cite{Gasperini})
but on the particular configuration itself.
In some sense, our idea is that despite the key role
played by the cosmological constant at a more
fundamental level, the
mechanism that causes the universe to become classical
(at large scales) might be related closer to
the nucleation energy than to $\Lambda$.
The formal advantage of any approach
based on real tunneling configurations
is that the concept of energy ($E$) and
space ($\Sigma$) is associated with a certain
Euclidean solutions of the field equations.
\medskip

So far we have laid emphasis on nucleation scenarios.
A similar procedure may also be tried within a
wormhole-based model, although in a more
formal way since the significance of $\Sigma$ and
$E$ for the Euclidean configurations representing the
bunch of universes is not as transparent as
in the Hartle-Hawking framewok. However, as
already mentioned, it might be that some
quantum cosmological modification of Coleman's
appoach looks formally quite similar.
The modification of the standard procedures implied by our
suggestion, if it is taken over to the
wormhole model, may be illustrated by Coleman's proposal
that the value of certain
coupling constants (''constants of nature'', they
are contained in our $\gamma$) may be determined
by minimizing the effective action $I_\Lambda(\gamma)$ at
small constant $\Lambda\downarrow 0$
\cite{Coleman2}. In our approach, this is replaced by
minimizing $I_E(\alpha)$ (the coupling constants now being
contained in $\alpha$) at large constant
$E\rightarrow \infty$. As far as the dimension is
concerned, our approach (like Gasperini's
\cite{Gasperini}) will predict
$n=4$ as the lowest possible value,
whereas minimization at constant $\Lambda$ would
lead to $n\rightarrow\infty$.
It is not clear conceptually, however, whether the limit
$E\rightarrow\infty$ is to be understood as the
prediction that space is (tends to be) flat,
and whether at least the topoloy and dimension
of $\Sigma$ have some physical significance apart from
the issue of nucleation.
(Also, it is not clear to what extent $E$ and $\Sigma$
refer to ''our'' universe, or to the bunch of universes
that exist in the wormhole-based models).
The suggestion we made might be reasonable only for the early
universe which is about to nucleate. In any case, as soon as
the size of the universe exceeds the wormhole scale,
Coleman's original theory should begin to apply, possibly
somehow ''decoupled'' from the large-scale phenomena such
as the dimension and topology of space.
Also, whether our procedure can effectively be made
equivalent to the standard wormhole model as far
as the late universe is concerned (e.g. whether the
quantities $E$ and $\Sigma$ can be given a conceptually
convincing role in a path integral), must be left open.
We will find a certain hint for further speculations in
Section 7 (equation (\ref{zsfdssdg})).
\medskip

Due to our lack of knowlegde about the details of the
underlying nucleation or large-scale decoherence
mechanism, the probabilistic
interpretation might apply only approximately.
In other words, the process of ''freezing out''
a particular $(\Sigma,h,\phi)$ (to play the role of the
classical initial data)
can reasonably be expected to select some
configuration whose action is small but maybe non-minimal.
In a model containing wormholes,
such a mismatch between propabilistic law and actual
dynamics may also be due to an effect induced
by the existence of a wormhole scale (whose
magnitude is unknown, cf. Ref. \cite{Coleman2}).
One may imagine for example some mechanism
that ''freezes out'' a classical scale
factor just above the wormhole scale, and hence at
an energy scale $E$ that we do not know exactly.
The suppression of a subsequent tunneling (dimensional
transition) into a lower dimensional state
would of course have to be explained in such a
scenario.
We shall take into account this possibility when
interpreting our numerical results in Section 5.
\medskip

In the whole of this paper we ignore higher loop
contributions. It is likely that more appropriate (and
accurate) results can be achieved by redefining the
nucleation probabilities as $P\sim A \exp(-I)$, where $A$
is the one-loop prefactor arising from fluctuations
around the Euclidean configuration
(see e.g. Refs. \cite{HawkingQG} and \cite{GibbonsPerry}).
Hence, our results certainly become unreliable as $E$ gets
very small. However, including these corrections,
we do not expect the principle structure of
the analysis and its conceptuals difficulties to change
dramatically.
Also, we do not consider the approach to quantum cosmology
due to Vilenkin and Linde, based on the so-called
''tunneling wave function'' of the universe
\cite{Vilenkin1},
\cite{Vilenkin2}--\cite{Vilenkin3}).
In a sense, the Euclidean action enters this formalism with
different sign, so that one ends up with probabilities
$P\sim\exp(I)$. Naively replacing the minimization
by maximization of $I$ in our framework, one would obtain
arbitrarily high nucleation dimensions to be favoured.
Let us just mention that this could be avoided
within our proposal by an anomalous behaviour
of the gravitational constant for $n\rightarrow\infty$
(cf. equation (\ref{gravconst}) below and the discussion
of the $\kappa_n$ at the beginning of Section 6).
\medskip

Moreover, we will neither specify the particular matter
content of the fundamental theory, nor of the
effective action in the wormhole-based models.
However, as far as nucleation is concerned,
the initial configuration $(\Sigma,h,\phi)$
carries some (effective) non-zero cosmological
constant $\Lambda$. In the Hartle-Hawking
formalism without wormholes, $\Lambda$ may survive for some
time period
(during which it drives an inflationary expansion) until
it eventually ''decays'', whereas in a model including
wormholes it should rapidly fall to zero, once the size of
the universe exceeds the wormhole scale
\cite{Coleman2} (although
the details of this process are unknown).
Whether the universe will actually undergo
an inflationary phase after nucleation will certainly
be affected by the answer to these (and related)
fundamental questions,
as well as by the field content and matter couplings.
Moreover, in the case the universe nucleates
at a dimension larger than four, and the initial hypersurface
has the product structure
$\Sigma = {\cal N}_{n-4}\times {\cal N}_3$, it
is conceivable that the field couplings
give rise to some Kaluza-Klein mechanism
that captures the factor
${\cal M}_{n-4}$ at unobservably small sizes
(see e.g. \cite{FreundRubin} and \cite{Volovich}).
In this case,
space-time appears to be four-dimensional although at a
more fundamental level it is of dimension $n$
(see e.g. Refs. \cite{BailinLove} and \cite{Duff}).
If such matter couplings are absent, the primordial
(''frozen out'') cosmological constant will ''decay'' or
be driven to zero on account of the wormhole mechanism, and
eventually all dimensions blow up. We will keep in mind
that we cannot exclude any of these possibilities at the
principal level of our considerations.
\medskip

Returning to the technical aspects of our suggestion,
we have to note that there is one unknown fundamental
quantity remaining, namely the
$n$-dimensional gravitational constant $G_n$, as
showing up in (\ref{2.1}). We exclude it from effectively
depending on other vaiables.
(Since in the wormhole-based models it appears to be
bounded away from zero anyway, our assumption to treat
it as a true constant does not seem to be an
oversimplification; cf. Refs.
\cite{Preskill} and \cite{HawkingW}).
{}From dimensional arguments,
we know that $G_n\sim m_P^{2-n}$, but we do not now
anything about the prefactor if $n\neq 4$. For
later convenience, let us write
\begin{equation}
G_n = \left( \frac{\kappa_n}{m_P}\right)^{n-2}\, ,
\label{gravconst}
\end{equation}
where the $\kappa_n$ are dimensionless numbers,
and $\kappa_4=1$.
Sometimes all $\kappa_n$ are set equal to $1$
(e.g. in Ref. \cite{Gasperini}).
However, this need not be the correct choice.
When relative probabilities of configurations in
different dimensions are compared, one would have
to know these numbers. Put otherwise, one
can try to set bounds on them by means of
physical predictions concerning our universe. However,
it appears natural to assume that the $\kappa_n$ are
of the order 1,
and to check the sensitivity of predictions against
small modifications thereof.
This is what we will do in Section 6,
but we should keep in mind that a full understanding
of a genuinely multiple-dimensional theory, and a
quantum cosmological explanation why we observe a
four-dimensional space-time relies on these numbers
as well.
\medskip

In the remaining Sections we will specialize to a simple
model in which our suggestion can be tested. As a model for
$I$ (whether it is intepreted as fundamental or effective)
we consider pure gravity, matter
just being represented effectively by a (non-fixed)
cosmological constant $\Lambda$.
(This corresponds to Coleman's original scenario
\cite{Coleman2}).
Moreover, we will not consider all solutions
to the Euclidean Einstein equations (i.e. all Einstein
metrics) admitting a symmetric hypersurface
$\Sigma$, but only products of (round) spheres of
arbitrary dimension. Since not much is known
about the set of all Einstein spaces admitting a
symmetric hypersurface (cf. \cite{GibbonsHartle}),
and in view of the belief that this set is probably
quite small \cite{MKriele}, the restriction to
products of spheres may turn to be
a viable approximation of the real situation.
Within this framework, we can carry out the
analysis of relative probabilities and ask for
the favoured dimension and topology of the
universe. Due to the product structure of the manifolds
we consider, we will obtain the principal possibility
to include the question whether our universe
is of a Kaluza-Klein type, i.e. with several
dimensions compactfied to unobservably small size.
We will find more or less definite answers to
these questions.
However, when dealing with numbers, we should not
forget that higher loop corrections
as well as sophisticated matter sectors
could modify our results.
\medskip

\section{Action on products of spheres}
\setcounter{equation}{0}

In this Section we consider manifolds which are, topologically,
products of spheres, i.e.
\begin{equation}
{\cal M} = {\bf S}^{n_1} \times {\bf S}^{n_2} \times \dots
\times {\bf S}^{n_m}\, ,
\label{pr1}
\end{equation}
and denote the total dimension of ${\cal M}$ by
\begin{equation}
n =\sum_{B=1}^m n_B\, ,
\label{dim}
\end{equation}
$m$ being the number of factors. Any of the spheres
${\bf S}^{n_B}$ shall carry the standard (round) metric
with radius $a_B$, which implies that its curvature scalar
is given by $R_B = n_B(n_B -1) /a_B^2$.
\medskip

Our next assumption is that the total metric thus defined
on ${\cal M}$ satisfies the Euclidean Einstein equations
(\ref{2.2}) with cosmological constant $\Lambda>0$
(and $T^{\rm \,else}_{\mu\nu}=0$).
Here, some trivial cases are possible: If $n=1$ or $n=2$,
the Einstein equations lead to $\Lambda=0$. If $n>2$ and
at least {\it one} of the spheres is one-dimensional, the
Einstein equations imply $n_B=1$ for {\it all} $B=1,\dots m$,
and $\Lambda=0$. We will not consider these geometries, and
restrict the sequence of dimensions $\{n_1,\dots n_m\}$ to the
generic case $n>2$ and $n_B>1$ for all $B$. Hence, the lowest
possible dimension is $n=3$, and it is realized only by
${\cal M} = {\bf S}^3$ (this configuration will later on
be excluded by a mechanism similar
to the one proposed by Gasperini \cite{Gasperini}).
The next possible dimension,
$n=4$, is realized by ${\cal M} = {\bf S}^4$ and
${\cal M} = {\bf S}^2\times {\bf S}^2$.
\medskip

In the generic case, as specified above, the Einstein equations
are equivalent to the system of equations
\begin{equation}
\frac{n_B-1}{a_B^2} = \frac{2\Lambda}{n-2}
\qquad B=1,\dots m\, .
\label{pr2}
\end{equation}
Hence, for any given $\Lambda>0$, the radii of the
factor-spheres are uniquely determined to be
\begin{equation}
a_B = \left(  \frac{(n_B-1)(n-2)}{2\Lambda}  \right)^{1/2}\qquad
B=1,\dots m\, .
\label{pr3}
\end{equation}
Any Einstein space constructed in this way has non-zero
curvature, and its volume is given by
\begin{equation}
{\cal V} = \left( \frac{n-2}{2\Lambda}\right)^{n/2}
\prod_{B=1}^m v_{n_B} (n_B-1)^{n_B/2} \, ,
\label{pr4}
\end{equation}
where
\begin{equation}
v_p =
\frac{ 2 \pi^{(p+1)/2}}{\Gamma(\frac{p+1}{2})}
\label{sph12.5}
\end{equation}
is the volume of the unit $p$-sphere.
The value of the Euclidean action (\ref{2.1}), when such a
solution is inserted, turns out to be (using
$R = 2n\Lambda/(n-2)$ for the total curvature scalar,
which follows from contracting (\ref{2.2}))
\begin{equation}
I = -\,\frac{{\cal V}}{8 \pi G_n} \frac{2 \Lambda}{n-2} =
-\,\frac{1}{8\pi G_n}
\left( \frac{n-2}{2\Lambda}\right)^{(n-2)/2}
\prod_{B=1}^m v_{n_B} (n_B-1)^{n_B/2} \, .
\label{pr4.1}
\end{equation}
For given $n$ and $\Lambda$, the minimum Euclidean action is
attained for the single-sphere ($m=1$) configuration
${\cal M}={\bf S}^n$, and this
provides, as already mentioned, a general lower bound for the
value of the Euclidean action, evaluated at
arbitrary compact solutions of the Euclidean Einstein
equations \cite{Besse}. The variables symbolically denoted by
$\gamma$ in Section 2 are now provided by
$\{n_1,\dots n_m\}$, and (\ref{pr4.1}) is just the expression
$I_\Lambda(\gamma)$.
\medskip

Any of our configurations $\cal M$ carries
hypersurfaces $\Sigma$ with zero extrinsic
curvature (i.e. symmetric hypersurfaces),
at which the universe can nucleate.
Since our configurations ${\cal M}$ are
products of spheres, the possible hypersurfaces $\Sigma$
are provided by the equators of the factor-spheres.
Let $\Sigma_A$ denote the
hypersurface of $\cal M$ which corresponds to the equator
of the $A$-th sphere ${\bf S}^{n_A}$ (it is defined as the
set of all points $(x_1,\dots,x_A,\dots,x_m)$, where
$x_A$ is on the equator of the $A$-th sphere, while all
other $x_B$ are points on the $B$-th sphere without further
restriction). This means that we assume a situation
analogous to (\ref{signch}) in the $A$-th factor-sphere.
In terms of classical signature change, one
${\bf S}^{n_A}$ induces the emergence of a
Lorentzian time coordinate and a subsequent classical
evolution of the corresponding spacelike slices (which
are topologically products of
${\bf S}^{n_A-1}$ with the remaining unaffected spheres).
If we assume
$\Lambda$ to maintain its (non-zero) value
after nucleation (and henceforth be a constant),
the ${\bf S}^{n_A}$-part
of $\Sigma_A$ will inflate while all
other factor-spheres remain ''small''. (In the language
of Kaluza-Klein theories, one would state that they
remain ''compactified'').
However, since $\Lambda$ is only an effective
phenomenon, it might rapidly decay as well,
the details depending on the particular theory that
is approximated by our simplified model, and
we shall not specify them.
\medskip

Thus, assuming the $A$-th sphere of ${\cal M}$ to
induce a change of metric signature,
we arrive at a ''real tunneling geometry''
\begin{equation}
{\cal M}_{\rm real\,\, tunneling} =
{\bf S}^{n_1} \times \dots
\times {\bf K}^{n_A}
\times \dots\times{\bf S}^{n_m}\, .
\label{realtunn}
\end{equation}
The Euclidean part of this space is just the half of
(\ref{pr1}), and the matching to the classical Lorentzian
universe is along $\Sigma_A$.
\medskip

Having specified the family of pairs $({\cal M},\Sigma)$
that may occur in this simplified version of our
proposal, we begin carrying out the analysis
described in the foregoing Section by computing the
nucleation energy $E$. Once $({\cal M},\Sigma)$
(i.e. $\{n_1,\dots n_m\}$ and $A$) are fixed,
the effective non-gravitational energy generated at
$\Sigma_A$ is given by
\begin{equation}
E = \frac{1}{8\pi G_n}\, {\cal V}_{\Sigma_A}\, \Lambda\, ,
\label{E1}
\end{equation}
where
\begin{equation}
{\cal V}_{\Sigma_A} =
\frac{v_{n_A - 1}}{v_{n_A} a_A}\,\,{\cal V} =
    \left(\frac{n-2}{2\Lambda}\right)^{(n-1)/2}
\frac{v_{n_A -1}}{v_{n_A} (n_A - 1)^{1/2}}
    \prod_{B=1}^m v_{n_B} (n_B-1)^{n_B/2}
\label{pr5}
\end{equation}
is the volume of $\Sigma_A$ as an $(n_A-1)$-manifold. As a
consequence,
\begin{equation}
E
= \frac{1}{8\pi G_n}\, \frac{v_{n_A-1}}{v_{n_A} a_A}\, {\cal V}\,\Lambda
=-\, \frac{v_{n_A-1} \,(n-2) }{2\, v_{n_A} a_A}\,\, I.
\label{tzuwq}
\end{equation}
When expressed entirely in terms of $\Lambda$ (using (\ref{pr3})
and (\ref{pr4})),
we obtain
\begin{equation}
E =
\frac{1}{16\pi G_n}\,\,
\frac{v_{n_A-1} \,(n-2) }{ v_{n_A} (n_A-1)^{1/2}}\,\,
\left( \frac{n-2}{2\Lambda}\right)^{(n-3)/2}
\prod_{B=1}^m v_{n_B} (n_B-1)^{n_B/2} \, .
\label{pr9}
\end{equation}
For any given configuration (with $n\neq 3$)
this provides a one-to-one correspondence
between the cosmological constant and the nucleation
energy. The set of variables denoted by $\alpha$ in
the preceding
Sections is now given by $\{n_1,\dots n_m\}$ together
with $A$.
The Euclidean action may thus likewise be expressed
in terms of the variables $(\Lambda,\gamma)$ or
$(E,\alpha)$.
\medskip

We note that for $n=3$, the energy is independent of $\Lambda$,
and is given by $E = 1/(6 G_3)$. Hence, if $E$ plays the role of
a quantity whose value is generic and only dynamically
governed by a probability law, it appears unlikely that
the $n=3$ case will contribute to the fluctuations between
(or succession of) our various configurations $\cal M$.
Hence, in this sence, the minimum dimensionality of space-time
is predicted to be $4$. This is quite similar to the reasoning
of Gasperini \cite{Gasperini}.
\medskip

If $n>3$, the relation (\ref{pr9}) can be used to express
$\Lambda$ in terms of $E$.
Inserting the result into
(\ref{pr4.1}), we find the action expressed entirely in terms
of the energy $E$, the dimensions $\{n_1,\dots,n_m\}$
(characterizing the topology of
$\cal M$) and the number $A$ of the sphere which
induces the transition to a classical universe.
Using (\ref{gravconst}) in addition,
we arrive at the expression
\begin{equation}
I =
- \left(  \frac{2\, \kappa_n}{n-2}\,\frac{E}{m_P}\,
\frac{v_{n_A} (n_A-1)^{1/2}}{v_{n_A-1}} \right)^\frac{n-2}{n-3}
\left( \frac{1}{8\pi}
       \prod_{B=1}^m v_{n_B} (n_B-1)^{n_B/2}
\right)^{-\,\frac{1}{n-3}} \, ,
\label{pr10}
\end{equation}
which corresponds to what
was denoted by $I_E(\alpha)$ in Section 3.
It is the starting point for our quantitative
analysis.
\medskip

\section{Favoured topologies at fixed dimension}
\setcounter{equation}{0}

According to our proposal, the first step is to
analyze the values of the action for constant $E$.
Here we encounter the fact that $I$ from (\ref{pr10})
contains the unkown numbers $\kappa_n$.
However, if $n$ is fixed in addition,
we can at least distinguish between different topologies.
Hence, let us postpone the question which dimension
minimizes $I$ at a given value of $E$ to the next Section,
and suppose $n$ to be fixed as well as $E$.
Thus the question which topology is favoured if the
universe nucleates as an $n$-dimensional space-time
becomes a well-posed problem.
\medskip

In order to get a more handsome quantity, we omit
the irrelevant factors in $I$ containing
only $n$ and $E$, and raise to the power $3-n$ (which is
assumed to be negative from now on). The most probable
configuration is the one minimizing
\begin{equation}
 F  =
 \left(
\frac{v_{n_A-1}}{v_{n_A} (n_A-1)^{1/2}} \right)^{n-2}
\prod_{B=1}^m v_{n_B} (n_B-1)^{n_B/2}\,.
\label{maxi}
\end{equation}
The complete Euclidean action is given by
\begin{equation}
I =
- \left(\frac{8\pi} F \right)^{1/(n-3)}
\left(  \frac{2\, \kappa_n}{n-2}\,\frac{E}{m_P}
\right)^{(n-2)/(n-3)}  \, .
\label{pr1011}
\end{equation}
Minimizing $ F $ is easily done for small values of $n$ by
explicitly inserting the candidate configurations. In order
to have a convenient notation, we will denote the $A$-th
sphere (the one that undergoes a signature change) by
a tilde, hence write
\begin{equation}
{\cal M} = {\bf S}^{n_1}\times\dots
\times \widetilde{{\bf S}^{n_A}}\times\dots{\bf S}^{n_m}\, .
\label{config}
\end{equation}
For $n=4$ and $n=5$ we find
\begin{eqnarray}
 F (\widetilde{{\bf S}^4}) &=& \frac{9\pi^2}{2}\approx 44.4132
\label{list1}\\
 F ({\bf S}^2\times\widetilde{{\bf S}^2}) &=&
4\pi^2\approx 39.4784
\label{list2}\\
 F (\widetilde{{\bf S}^5}) &=&\frac{2048}{27}\approx 75.8519
\label{list3}\\
 F (\widetilde{{\bf S}^2}\times{\bf S}^3) &=& 2\sqrt{2}\pi^3
\approx 87.699
\label{list4}\\
 F ({\bf S}^2\times\widetilde{{\bf S}^3}) &=& 64\, .
\label{list5}
\end{eqnarray}
Hence, in $4$ dimensions,
${\bf S}^2\times\widetilde{{\bf S}^2}$ is favoured
over $\widetilde{{\bf S}^4}$, and in $5$ dimensions
${\bf S}^2\times\widetilde{{\bf S}^3}$ is the winner.
Let us write down a list of the configurations ${\cal K}_n$
minimizing $F$ for $n=4,\dots 15$
\begin{eqnarray}
{\cal K}_4 &=& {\bf S}^2\times\widetilde{{\bf S}^2}
\label{lll4}\\
{\cal K}_5 &=& {\bf S}^2\times\widetilde{{\bf S}^3}
\label{lll5}\\
{\cal K}_6 &=& {\bf S}^2\times\widetilde{{\bf S}^4}
\label{lll6}\\
{\cal K}_7 &=&
{\bf S}^2\times{\bf S}^2\times\widetilde{{\bf S}^3}
\label{lll7}\\
{\cal K}_8 &=&
{\bf S}^2\times{\bf S}^2\times\widetilde{{\bf S}^4}
\label{lll8}\\
{\cal K}_9 &=&
{\bf S}^2\times{\bf S}^2\times\widetilde{{\bf S}^5}
\label{lll9}\\
{\cal K}_{10} &=&
{\bf S}^2\times{\bf S}^2\times
{\bf S}^2\times\widetilde{{\bf S}^4}
\label{lll10}\\
{\cal K}_{11} &=&
{\bf S}^2\times{\bf S}^2\times
{\bf S}^2\times\widetilde{{\bf S}^5}
\label{lll11}\\
{\cal K}_{12} &=&
{\bf S}^2\times{\bf S}^2\times{\bf S}^2\times
{\bf S}^2\times\widetilde{{\bf S}^4}
\label{lll12}\\
{\cal K}_{13} &=&
{\bf S}^2\times{\bf S}^2\times{\bf S}^2\times
{\bf S}^2\times\widetilde{{\bf S}^5}
\label{lll13}\\
{\cal K}_{14} &=&
{\bf S}^2\times{\bf S}^2\times{\bf S}^2\times
{\bf S}^2\times\widetilde{{\bf S}^6}
\label{lll14}\\
{\cal K}_{15} &=&
{\bf S}^2\times{\bf S}^2\times{\bf S}^2\times
{\bf S}^2\times{\bf S}^2\times\widetilde{{\bf S}^5}\, .
\label{lll15}
\end{eqnarray}
Hence, there is a clear tendency for $F$ to be minimized
by products of several ${\bf S}^2$ with a higher dimensional
${\bf S}^p$.
Moreover, in all configurations, the signature changing
sphere is the one with the highest dimension. This is
valid for all $n$ (it is a trivial consequence of the
numerics of $v_p$).
For the dimensions $(16,\dots 42)$ the
according configurations are products of two-spheres
with an $\widetilde{{\bf S}^p}$, where $p$ takes the values
$(6,5,6,7,6,7,6,7,6,7,8,7,8,7,8,7,8,9,8,9,8,9,8,9,8,9,10)$.
Hence, the increase of $p$ with $n$ is rather modest,
and $n=12$ is the highest
dimension in which a factor sphere ${\bf S}^4$ exists.
\medskip

Let us analyze what happens asymptotically for large $n$.
The configuration minimizing $F$ is a $q$-fold product of
two-spheres with an $\widetilde{{\bf S}^p}$, hence $m=q+1$
and $2 q + p = n$. Evaluated at these numbers, $F$
becomes
\begin{equation}
F = (4\pi)^{(n-p)/2}
\left(\frac{v_{p-1}}{v_p\,(p-1)^{1/2}}\right)^{n-2}
v_p\,(p-1)^{p/2}\, .
\label{asymp1}
\end{equation}
Treating $p$ as a continuous variable, this expression
can be differentiated with respect to $p$. One
thereby encounters the so-called logarithmic derivative
of the Gamma-function whose expansion for large arguments
is
\begin{equation}
\Omega(x)\equiv
\frac{\Gamma'[x]}{\Gamma[x]} =
- \ln\left(\frac{1}{x}\right) -\,\frac{1}{2x}
-\,\frac{1}{12 x^2} +
O\left(\frac{1}{x^4}\right).
\label{loggamma}
\end{equation}
Since $\partial F/\partial p = 0 $ reduces to an
expression of first order in $n$, it can be solved
to give
\begin{equation}
n = \frac{p + 2 + (p-1) \left(
\ln(p-1) -2 \ln 2 +2\,\Omega\left(\frac{p}{2}\right)
- 3\, \Omega\left(\frac{p+1}{2}\right)
           \right)}{1 +
(p-1) \left(
\Omega\left(\frac{p}{2}\right)
- \Omega\left(\frac{p+1}{2}\right) \right)}\, .
\label{hjjkwx}
\end{equation}
Insertion of the
asymptotic behaviour (\ref{loggamma}) yields
\begin{equation}
n = (1 - \ln 2) (2p^2-4p)+
\frac{17}{3}-3\ln 2 +
O\left(\frac{1}{p^2}\right).
\label{hcbs}
\end{equation}
Inverting the series, and  neglecting contributions which
vanish as $n\rightarrow\infty$,
we find the dimension of the signature
changing sphere in ${\cal K}_n$ for large $n$ to be
\begin{equation}
p\approx \beta \sqrt{n} + 1
\label{zcg}
\end{equation}
with $\beta=(2(1-\ln 2))^{-1/2}\approx 1.2765$.
This explains the slow increase of the factor
spheres' dimensions we observed before. The
error in (\ref{zcg}) will, by the way, be of order $1$
even for $n\rightarrow \infty$, because both $p$ and
$n$ are integers (a fact that is ignored by the
continuous approximation).
\medskip

In order to know the approximate value of $F$ at these
configurations, we use Stirling's formula for the
Gamma-function to obtain after a lengthy computation
\begin{equation}
F({\cal K}_n) =
2^{(n+2)/2}\, \pi\,
\exp\left(\frac{\sqrt{n}}{2\beta} \, \right)
\left( 1 + O\left(\frac{1}{\sqrt{n}}\right)
\right).
\label{hcksnhj}
\end{equation}
For the low dimensions this formula is of course not
very accurate but displays the qualitative behaviour quite
well. If $n$ varies from 4 to 15 (corresponding to
the list (\ref{lll4})--(\ref{lll15})), the
relative deviation from the exact value is between 12\%
and 29 \%.
\medskip

For comparison we also write down a similar
expansion for $F$ evaluated on single-sphere
($m=1$) configurations with large dimension,
\begin{equation}
F({\bf S}^n) =
2^{3/2}\,\pi\,
\exp\left(\frac{n}{2}-\frac{1}{4}\right)
\left( 1 + O\left(\frac{1}{n}\right)
\right).
\label{kjcbdf}
\end{equation}
The dominant contributions to the exponential
increase
\begin{equation}
F({\cal K}_n)\sim 2^{n/2}\, ,\qquad\qquad
F({\bf S}^n)\sim e^{n/2}
\label{asc}
\end{equation}
show how the products ${\cal K}_n$ are favoured over
the single-spheres ${\bf S}^n$.
\medskip

Do these results already give rise to predictions?
We will confirm in Section 7 that
-- as a consequence of our proposal --
the small dimensions are favoured over the large ones
when the universe nucleates.
Thus, if the minimization of $I$ is applied
straightforwardly, we would predict $n=4$ and the global
topology of the spatial sections of the universe to be
the same as the signature
changing hypersurface $\Sigma$ of ${\cal K}_4$,
hence ${\bf S}^1\times{\bf S}^2$.
In other words, the universe is of
Kantowski-Sachs type,
a result that might be not very appealing
(although consistent with observations).
However, we must keep in mind that
we have not inserted a particular model for the matter
sector (except for an effective cosmological constant).
This leaves some freedom for an alternative
scenario: As we mentioned in Section 3 the
mechanism selecting a classical universe might
pick a configuration with large (but possibly not the
largest) probability. In other words,
it might drive the universe classical
''before'' the absolute minimum of $n$
is reached. Furthermore, the wormhole scale
(which we do not know) provides some uncertainty
about these things.
Hence, if the selected configuration is not
${\cal K}_4$ but some other ${\cal K}_n$ (with small
$n$, which we may reasonably assume), and if its
nucleation is accompagnied with a subsequent
Kaluza-Klein mechanism
\cite{BailinLove} that captures some of the
scale factors, we might end up with a space-time
containing unobservably small internal spaces.
The most appealing candidates in our list
(\ref{lll4})--(\ref{lll15})
of minimizing configurations ${\cal K}_n$ are of course
those containing a factor $\widetilde{{\bf S}^4}$. One of the
four dimensions is
absorbed by the emergence of Lorentzian time, and the
according equator ${\bf S}^3$ represents the observable
space-like slices of our universe. This mechanism
is possible in $n=6$, $8$, $10$ and $12$ dimensions, the
internal spaces being products of two-spheres.
\medskip

We should add that including a more sophisticated matter
sector in the action $I$ could substantially
modify these results
(cf. Ref. \cite{FreundRubin} for supergavity,
Ref. \cite{Volovich} for six-dimensional
Einstein-Maxwell theory and
Ref. \cite{Myers2} for an axion field), and it is
presumably easily to find models in which
$\Sigma={\bf S}^3$ is the most probable initial slice,
even without Kaluza-Klein effect.
On the other hand, if
the possibility of wormholes is admitted,
and the non-gravitational degrees of freedom
do not give rise to vacuum energies other than those
represented by $\Lambda$, the model considered here
should apply quite well. In the case
such a version of a wormhole-based model
applies for energies $E$ substantially larger than the
wormhole scale, the prediction for the spatial slices is
${\bf S}^1\times{\bf S}^2$.
\medskip

Thus, even the first step in our procedure
(together with a slight anticipation of the
following) has revealed
non-trivial statements about the global topology
of space-time. The next -- somewhat intermediate --
step will be to examine the value of the action at
constant $E$ but variable $n$.
\medskip

\section{Favoured dimension at fixed energy}
\setcounter{equation}{0}

Turning to the question which dimension is favoured at
a given nucleation energy $E$, we run inevitably into
the problem that the numbers $\kappa_n$ in the
$n$-dimensional gravitational constant are unknown (except
for $\kappa_4 = 1$). In lack of a better knowledge about
this, we will base our arguments on the expectation
that $\kappa_n$ does not deviate much from unity.
Let us begin however leaving $\kappa_n$ unconstrained.
\medskip

The results of the foregoing Section amount to associate
with each value of $n$ a Euclidean configuration
${\cal K}_n$, minimizing the quantity $F$ from (\ref{maxi}).
Thus the question for the favoured dimension is traced back
to the minimization of $I_E({\cal K}_n)$ at constant $E$.
Using the form (\ref{pr1011}) and the
asymptotic behaviour (\ref{hcksnhj}),
we note that $F({\cal K}_n)^{-1/(n-3)}\rightarrow 1/\sqrt{2}$
as $n\rightarrow\infty$, and
\begin{equation}
I_E({\cal K}_n) =  -\sqrt{2}
\left(\frac{\kappa_n}{n-2}\right)^{(n-2)/(n-3)}
\,\frac{E}{m_P}
\left( 1 + O\left(\frac{1}{n}\right)\right).
\label{ugfewjkrt}
\end{equation}
This shows the extent to which the whole procedure is
sensitive to the values of $\kappa_n$. If
$\kappa_n/n \rightarrow 0$ as $n\rightarrow\infty$,
the action will approach zero from below, and will
definitely admit a finite value of $n$ where it
attains its minimum. If $\kappa_n/ n $ blows up,
$I$ will tend towards $-\infty$, thus having no
minimum. In what follows we assume the former case to
be true. (Let us just mention that in the case
$\kappa_n/n$ blows up as $n\rightarrow\infty$,
the action will attain a maximum at some finite $n$.
Such a modification could account for the reversed
sign the action enters the tunneling proposal
approach by Vilenkin and Linde \cite{Vilenkin1},
\cite{Vilenkin2}--\cite{Vilenkin3},
where $P\sim\exp(I)$).
\medskip

Let us return to the exact expressions and ask what happens
if $E$ is very large. The relevant term in $I$ is
then provided by $(E/m_P)^{(n-2)/(n-3)}$, which is,
for $E>m_P$, a monotonically decreasing function
in $n$ that approaches $1$ at infinity. As long as
$\kappa_n/n$ decreases to zero for large $n$,
the minimizing dimension can be made smaller
by increasing $E$. Hence, if $E\gg m_P$, the action
is minimized at the lowest possible dimension,
$n=4$. For such a large value of $E$ one has,
clearly, $I_E({\cal K}_4)>I_E({\cal K}_n)$ for all
$n\geq 5$. Decreasing $E$, things can change only if
at some threshold $E=E_4$, the action of ${\cal K}_4$
becomes equal to the action of some higher ${\cal K}_n$.
Note that, from the outset, it is not at all clear
whether such a threshold would occur at $n=5$ or
at some higher dimension. Again, this depends on the
constants $\kappa_n$ (and in particular on their
values for small $n$).
\medskip

A selection of energy values defined by such
identities is
\begin{eqnarray}
I_E({\cal K}_4)=I_E({\cal K}_5)\qquad {\rm at}\qquad
E &=& \frac{\pi^3}{108}\,
\kappa_5^3 \kappa_4^{-4}\,m_P\,,
\label{hghg1}\\
I_E({\cal K}_4)=I_E({\cal K}_6)\qquad {\rm at}\qquad
E &=& \frac{\sqrt{\pi}}{3\sqrt{6}}\,
\kappa_6^2 \kappa_4^{-3}\,m_P\,,
\label{hghg2}\\
I_E({\cal K}_5)=I_E({\cal K}_6)\qquad {\rm at}\qquad
E &=& \frac{432}{\pi^7}\,
\kappa_6^8 \kappa_5^{-9}\,m_P\, .
\label{hghg3}
\end{eqnarray}
This illustrates that one can easily create
various different situations by accordingly
adjusting the $\kappa_n$. This is of course not what we
are interested it. A reasonable condition on the
$\kappa_n$ is certainly provided by some sort of
monotonicity.
The result will in general be a monotonic
order of the favoured dimension as $E$ decreases.
In order not to overcomplicate things, let us
from now on make the usual choice
\begin{equation}
\kappa_n=1\qquad {\rm for\,\,all\,\,}n\,.
\label{kappa}
\end{equation}
Small modifications of these numbers will not change
any of our principal conclusions. (Recall that
the prediction of ${\cal K}_4$ as the
nucleation geometry at scales
$E\gg m_P$ is not affected thereof.)
\medskip

Having fixed the remaining freedom, the further analysis is
straightforward. The first ''transition'' occurs at
$I_E({\cal K}_4)=I_E({\cal K}_5)$, hence
\begin{equation}
E_4 = 2^{-2}\, 3^{-3}\, \pi^3\,m_P \approx 0.2871\,m_P\,.
\label{E0}
\end{equation}
Let us define in general $E_n$ by ${\cal K}_n$ and
${\cal K}_{n+1}$ having equal action. We easily find the
first few values
\begin{eqnarray}
E_5 &=& 2^4\, 3^3\, \pi^{-7}\, m_P
\approx 0.1430\, m_P\,,
\label{en5}\\
E_6 &=&
3^{12}\, 2^{-11}\, 5^{-15}\, \pi^{14}\,m_P
\approx 0.07757\,m_P\,,
\label{en6}\\
E_7 &=&
2^{37}\, 5^{25}\, 3^{-40}\, \pi^{-22} \,m_P
\approx 0.03892\,m_P\, .
\label{en7}
\end{eqnarray}
Thus we encounter a decreasing sequence of energy
scales. For $E>E_4$ the action is minimized by
${\cal K}_4$, hence the dimension $4$. If
$E_{n-1}>E>E_{n}$, the favoured configuration is
${\cal K}_{n}$, the according dimension is $n$.
\medskip

The asymptotic regime of small $E$ is governed by
the expansion (\ref{hcksnhj}). Again, an awkward
computations, which needs the expansion
\begin{equation}
\frac{F({\cal K}_n)}{F({\cal K}_{n+1})}=
\frac{1}{\sqrt{2}}\,\exp\left(
-\,\frac{1}{4\beta\sqrt{n}}\right)
\left( 1 + O\left(\frac{1}{n^{3/2}}\right)\right)
\label{ckhv}
\end{equation}
as as intermediate step, reveals for large $n$
\begin{equation}
\frac{E_n}{m_P} = \frac{n}{2^{3/2}}\,
\exp\left( - n +\frac{\sqrt{n}}{4\beta}+
\frac{5}{2}\right)
\left( 1 + O\left(\frac{1}{\sqrt{n}}\right)
\right).
\label{hbxs}
\end{equation}
Note that the level spacing
$\Delta E \equiv E_{n-1}-E_{n}$ is of the order
of $E$ itself, $\Delta E \approx E$.
Inverting this expression up to the first two orders,
it follows that the favoured dimension for
$E\ll m_P$ is
\begin{equation}
n \approx \ln\left(\frac{m_P}{E}\right)
+ \frac{1}{4\beta}
\left(\ln\left(\frac{m_P}{E}\right)\right)^{1/2}\, .
\label{whdg}
\end{equation}
Although this formula is probably going to be
modified by higher loop corrections, it nicely
display what one means by a higher-dimensional
phase in the early universe. In absence of a
decoherence mechanism as very small energies,
one could even say that all of these dimensions
contribute and talk about a multiple-dimensional
state. This can be viewed as a quantum version of
the idea that the universe ''begins'' as a
cold and small (though higher-dimensional) one.
(The nucleation volume is in fact very small
for large $n$, whereas
the nucleation radius is large, see equations
(\ref{rst1}) and (\ref{rst2}) below).
\medskip

As an alternative method, one could have improved
the expansion (\ref{hcksnhj}) by including the
next order, and differentated $I$ directly with respect
to $n$ in order to find the minimizing dimension.
In order to further characterize the
configurations ${\cal K}_n$, one may derive
asymptotic expressions for their total action,
the nucleation volume
${\cal V}_\Sigma\equiv {\cal V}_A$ and the
total $n$-volume ${\cal V}$,
as well as for the radius $a_A$ of the nucleating
factor $\widetilde{{\bf S}^p}$ and the
according value of the cosmological constant $\Lambda$.
In order not to present too much technical details,
we just write down qualitatively the leading
behaviour for
large $n$ (small $E$)
\begin{eqnarray}
- I &\sim&
{\cal V}_A\sim {\cal V} \sim e^{-n}\sim E\label{rst1}\\
a_A &\sim& n^{1/4}\sim (- \ln E)^{1/4}\label{rst2}\\
\Lambda &\sim& n \sim -\ln E\,. \label{rst3}
\end{eqnarray}
Note that due to the bunch of ${\bf S}^2$-factor-spheres
in the higher ${\cal K}_n$, the relation between the
cosmological constant and the nucleation radius is
not the standard one ($\Lambda\sim a_A{}^4$ instead of
$\Lambda\sim a_A{}^{-2}$, see also equation
(\ref{pr3})).
On the other side of the spectrum, for $E>E_4$, the
favoured configuration is ${\cal K}_4$, which implies
\begin{equation}
\Lambda=\pi^2 \left(\frac{m_P}{E}\right)^2 m_P^2
= a_A{}^{-2}\, .
\label{zgcjhvf}
\end{equation}
\medskip

\section{Favoured energy}
\setcounter{equation}{0}

Having exhibited the favoured configurations at fixed
$E$, the last step is to evaluate the values of the
action among these. (Asymptotically for
small $E$, we have done this already
in the preceding Section, see (\ref{ugfewjkrt}) and
(\ref{whdg})).
One might try to relate this step to the
question for the most probable energy the actual
classical universe will start with. However, the answer
will be $E=\infty$, and, as already stated,
we prefer to talk about some
mechanism that will drive the
universe to become classical
at a finite value of $E$. In Section 6 we found a
threshold $E_4$
above which the variables $\{n_1,\dots n_m\}$ and $A$
have settled and remain constant.
This threshold is of course a natural candidate for
the transition to a real universe.
However, since we neither understand the details of this
process nor have any knowledge of the wormhole
scale $E_{\rm wh}$, it is conceivable that nucleation
occurs only approximately at $E_4$. In the case
of $E<E_4$, it may not be the true minimizing configuration
$\Sigma={\bf S}^1\times{\bf S}^2$ (from ${\cal K}_4$)
that is ''frozen out'' but one that can be expected
to be among the next few on the list
(\ref{lll4})--(\ref{lll15}).
As already mentioned in Section 5, this provides the
possibility for a Kaluza-Klein scenario.
\medskip

Since the minimizing configurations at fixed $E$
(i.e. the functions $\alpha(E)$
in the notation of Section 3)
are already determined, there remains rather little
work to do. Let us denote by $I(E)$ the action at the
favoured configuration associated with the energy $E$,
i.e. the quantity we have called $I_E(\alpha(E))$
symbolically. For small $E$, (\ref{ugfewjkrt}) and
(\ref{whdg}) can be combined to give, to leading order,
\begin{equation}
I(E)\approx -\sqrt{2} \left(\ln\left(
\frac{m_P}{E}\right)\right)^{-1}
\frac{E}{m_P}\, ,
\label{3cec}
\end{equation}
which implies $\ln(-I) \sim \ln(E)-\ln(-\ln(E))$,
hence a behaviour quite close to linear ($I\sim -E$),
and in particular $I(0)=0$. Inside any interval
$E_{n-1}<E<E_n$ the favoured dimension is $n$, and the
action is given by (\ref{pr1011}), i.e.
\begin{equation}
I(E) = - k_n \, E^{(n-2)/(n-3)}
\label{hfdsag}
\end{equation}
where $k_n$ is a positive constant. This is a monotonically
decreasing function. Furthermore, from Section 6
we know that $I(E)$ is continuous, because is has
equal value for both adjacent configurations at the
boundaries $E_n$ of the intervals. Finally, for
$E>E_4$, we have dimension $4$ and $I(E)=- k\, E^2$.
To summarize, $I(E)$ is a continuous, monotonically
decreasing function, thus favours large $E$.
Thereby, the formal result of minimizing the
action, $E\rightarrow\infty$, is technically analogous
to $\Lambda\downarrow 0$ in Coleman's theory.
\medskip

This is the final (although heuristic)
step in our argumentation: The relative order that is
induced by the action on the set of ${\cal K}_n$ is from
small to large energies, hence from high to low
dimensions. At $E\approx E_4$, the large-scale variables
(like dimension and topology) become classical (whereas
the small-scale degrees of freedom might continue to
be quantum mechanical).
Hence, the energy scale $E\approx E_4$ is related to the
phase of nucleation,
while the scale $E\ll E_4$ corresponds to the genuine
multiple-dimensional quantum state.
\medskip

In the case our model is intended to represent a
wormhole-based scenario, one might ask whether
the framework we have developed is of some
significance for the late universe too. For $E>E_4$,
the minimizing configuration is ${\cal K}_4$. Due to
(\ref{zgcjhvf}), $\Lambda$ is predicted to approach
zero as $E$ blows up. The effective action for the late
universe (as long as ${\cal K}_4$ is actually selected to
become real) is given by
\begin{equation}
I({\cal K}_4) = -\,\frac{2}{\pi}
\left(\frac{E}{m_P}\right)^2
= - 2 \pi\,\frac{m_P^2}{\Lambda}\,.
\label{zsfdssdg}
\end{equation}
It can formally be inserted into Coleman's
measure for the wormhole parameters
(i.e. the variables $\Lambda$ depends on).
Since it differs from $I({\bf S}^4)$ only by a
constant prefactor, the familiar
$\delta(\Lambda)$-peak is produced. In terms of the
energy, this would result into the (naive) prediction
that $E$ is not just very large but actually infinite.
A possible implementation of our suggestion
in Coleman's model could be to consider $E$
as a function of the wormhole parameters, and to treat
the appearance of arbitrarily large universes
not just by a cutoff that is sent to infinity
\cite{Coleman2}, but by some cosmologically more
appealing regularization method.
Thus it would be a major challenge to modify or
rephrase the wormhole formalism such that is can
give rise to sensible predictions for global
cosmological variables (as e.g. $E$, if it can be
interpreted as the total matter energy contained
in the universe, and $\Sigma$, which may represent its
size as well as its large-scale topology).
The goal of such a theory would be to give
statements like $E\rightarrow\infty$
a significance as precise as the famous
''$\Lambda=0\,$'' in Coleman's work.
One possible question is
whether the ''nucleation picture''
becomes meaningless after nucleation (i.e. whether
the {\it large} universe may be understood in terms
of ''competing'' tunneling configurations as well).
The cosmological constant might in such a framework
just be a secondary quantity (the density associated with
$E$ and $\Sigma$, or, more general, the
constant $\Lambda$ appearing in (\ref{energy})).
The most reasonable point to make
contact between our framework and the standard
wormhole theory is provided by equation
(\ref{zsfdssdg}).
If the details of the decoherence mechanism or some
non-trivial matter sector induce nucleation around some
other configuration than ${\cal K}_4$, a modified action
$I\sim -E^{(n-2)/(n-3)}\sim$ $-\Lambda^{-(n-2)/2}$ takes
over the role of (\ref{zsfdssdg}).
\medskip

\section{Speculations}
\setcounter{equation}{0}

We would like to finish this paper by two comments.
The first one is devoted to the idea that
one could try to modify
the gravitational action by adding an appropriate
integral over $\partial{\cal M}$ to (\ref{2.1}),
as for example
\begin{equation}
m_P^{n-3} \int_{\partial {\cal M}}
d^{n-1} x\, \sqrt{h}
\left(c_1\, K^{ij}K_{ij} + c_2\, K^2\right).
\label{surface}
\end{equation}
Such terms might effectively be generated by higher
loop or higher order curvature contributions
(see Refs. \cite{Myers1}--\cite{MadsenBarrow}
for boundary terms in higher derivative gravity).
This can lead to a model in which the
stationary points of the action are only those
solutions of (\ref{2.2}) for which the
boundary data are time-symmetric (which implies in the
gravitational sector that the extrinsic curvature $K_{ij}$
vanishes). Given a manifold $\cal N$ with data
$(h,\phi)$ that are supposed to match a full
Euclidean solution $({\cal M},g,\Phi)$ at
$\partial {\cal M}={\cal N}$
(as is required in the Hartle-Hawking formalism), the
immediate consequence of $\delta I=0$
is that $({\cal M},g,\Phi)$
has to be a real tunneling configuration with
$\Sigma={\cal N}$.
Thus it is possible that for certain $({\cal N},h,\phi)$
(which are the arguments of the wave function $\psi$)
there is no stationary point at all.
In the case the data on $\cal N$ {\it do}
admit a properly matching Euclidean solution, the
dominant  contribution in the path
integral is given by $\exp(-I/2)$ with $I$ the
instanton action. This precisely reproduces the
nucleation amplitudes of the Hartle-Hawking
formalism. If, on
the other hand, the data on $\cal N$ do {\it not}
admit a properly matching Euclidean solution,
the path integral aquires contributions from various
non-stationary configurations, and are thus (leaving
aside questions of convergence) expected to be suppressed.
A peculiarity of such a formalism is that
the wave function does not necessarily oscillate when the
scale factors become large. (However, this need not
really be a drawback in an effective formalism).
In such a model it is not
necessary to ''associate'' the symmetric hypersurfaces
$\Sigma$ with the instantons by hand, because the
real tunneling configurations
play the role of dominant contributions at a
{\it fundamental} level.
\medskip

Our last comments concerns the approach suggested by
Gasperini \cite{Gasperini}, that the energy $U$
contained in the Euclidean $n$-geometry of an
instanton is conserved under dimensional transitions.
As already mentioned in Section 2,
it is not clear how $U$ relates
to an observed energy. Moreover, Gasperini's
framework involves only spheres, and it is not
obvious whether it can be generalized to instantons
or real tunneling configurations
with arbitrary topology. However, one may carry
over his idea to the simplified model we considered
in Sections 4 -- 7, by assuming that his definition
of $U$ applies for the $A$-th factor sphere, the
remaining spheres contributing their full volume,
just as contained in ${\cal V}_\Sigma$.
This means that (maybe contrary to the original spirit
of Gasperini's work) we use pairs
$({\cal M},\Sigma)$ in order to define $U$.
Thus, rescaling
$\widetilde{E}=2\pi U$, we set
\begin{equation}
\widetilde{E} =
\frac{1}{8\pi G_n}\, \frac{1}{a_A}\, {\cal V}\,\Lambda
=-\, \frac{n-2 }{2\, a_A}\,\, I \, ,
\label{pr11kjsd}
\end{equation}
hence
\begin{equation}
\widetilde{E} =
\frac{1}{16\pi G_n}\,\,
\frac{n-2}{ (n_A-1)^{1/2}}\,\,
\left( \frac{n-2}{2\Lambda}\right)^{(n-3)/2}
\prod_{B=1}^m v_{n_B} (n_B-1)^{n_B/2} \, .
\label{pr12cdhfds}
\end{equation}
As a consequence, the Euclidean action may be written as
\begin{equation}
I =
- \left(  \frac{2\, \kappa_n}{n-2}\,
\frac{\widetilde{E}}{m_P}\,
(n_A-1)^{1/2} \right)^\frac{n-2}{n-3}
\left( \frac{1}{8\pi}
       \prod_{B=1}^m v_{n_B} (n_B-1)^{n_B/2}
\right)^{-\,\frac{1}{n-3}}\,.
\label{pr13nkunlg}
\end{equation}
The according minimization procedure is to look for
minimizing configurations at fixed $\widetilde{E}$.
The analysis is very similar to the one we have
performed in Sections 4 -- 7, the essential difference
being that (\ref{pr13nkunlg}) favours the single-sphere
configurations ${\bf S}^n$ over the products of
spheres (hence over ${\cal K}_n$ as well). This is a
particularly
nice feature (it formally predicts not only $n=4$ as the
most probable dimension, but also the topology
of classical space-time to be ${\bf R}\times {\bf S}^3$).
The main disadvantage of this approach is in our
opinion -- apart from the questionable status of
$\widetilde{E}$ as observable quantity --
that it is not clear how it may be generalized to
arbitrary real tunneling configurations.
However, it may be considered as an alternative attempt,
sharing many technicalities with the one
based on the nucleation energy $E$.
\\
\\
\\
%{\Large {\bf Acknowledgments}}
%\medskip

\end{document}